\documentclass[twocolumn]{aastex7}

\usepackage{amsmath}
\usepackage{CJK}

\usepackage[normalem]{ulem}

\usepackage{xcolor}
\def\r{\textcolor{red}}
\def\b{\textcolor{blue}}
\def\c{\textcolor{cyan}}
\def\t{\textcolor{teal}}
\def\o{\textcolor{orange}}
\def\p{\textcolor{purple}}

\begin{document}

\begin{CJK*}{UTF8}{gbsn}

\title{Simulating the late stages of WD-BH/NS mergers: an origin for fast X-ray transients and GRBs with periodic modulations}

\author[0009-0001-7397-1727]{Jun-Ping Chen (陈军平)}
\affiliation{School of Physics and Astronomy, Sun Yat-Sen University, Zhuhai 519082, China} 
\affiliation{CSST Science Center for the Guangdong-Hongkong-Macau Greater Bay Area, Sun Yat-Sen University, Zhuhai 519082, China} \email{chenjp79@mail2.sysu.edu.cn}

\author[0000-0001-5012-2362]{Rong-Feng Shen (申荣锋)}
\affiliation{School of Physics and Astronomy, Sun Yat-Sen University, Zhuhai 519082, China}
\affiliation{CSST Science Center for the Guangdong-Hongkong-Macau Greater Bay Area, Sun Yat-Sen University, Zhuhai 519082, China} \email{shenrf3@mail.sysu.edu.cn}

\author[0000-0002-3525-791X]{Jin-Hong Chen (陈劲鸿)}
\affiliation{Department of Physics, University of Hong Kong, Pokfulam Road, Hong Kong, China}
\affiliation{The Hong Kong Institute for Astronomy and Astrophysics, University of Hong Kong, Hong Kong, China}
\affiliation{Shenzhen Institute of Research and Innovation, The University of Hong Kong, Shenzhen 518057, P. R. China}
\email{chenjh2@hku.hk} 

\author[0000-0003-3440-1526]{Wei-Hua Lei (雷卫华)}
\affiliation{Department of Astronomy, School of Physics, Huazhong University of Science and Technology, Luoyu Road 1037, Wuhan, 430074, China}
\email{leiwh@hust.edu.cn} 

\begin{abstract}
Recent studies indicate that mergers of a white dwarf (WD) with a neutron star (NS) or a stellar-mass black hole (BH) may be a potential progenitor channel for certain merger-kind, but long-duration $\gamma$-ray bursts (GRBs), e.g., GRBs 230307A and 211211A. The relatively large tidal disruption radius of the WD can result in non-negligible residual orbital eccentricity ($0 \lesssim e \lesssim 0.2$), causing episodic mass transfer, i.e., repeated partial disruptions (RPDs) of the WD. We perform smoothed-particle hydrodynamics simulations of RPDs in sixteen WD-BH/NS systems, capturing the subsequent mass transfer and accretion. The WD undergoes RPDs near the orbital periastron, modulating the ensuing accretion process, leading to variations of the accretion rate on the orbital period. Across all simulations, the peak accretion rates range from $4 \times10^{-4}$ to 0.16 $M_{\odot} \rm \ s^{-1}$, while the RPD duration spans from $\sim$ 10 s to an hour. More compact systems, i.e., those with a higher mass ratio (higher WD mass and lower accretor mass), tend to undergo fewer RPD cycles, resulting in shorter durations and higher accretion rates. If such events can launch relativistic jets, three categories of non-thermal X/$\gamma$-ray transients are predicted, in decreasing order of their mean accretion rates: (1) an X-ray transient with a simultaneous GRB, both lasting for $10^{1-2}$ s; (2) a longer X-ray transient lasting up to $10^{2-3}$ s that has a GRB appearing only at its later phase ; (3) an ultra-long X-ray transient lasting for $\sim 10^{3}$ s without a GRB. A generic feature of these transients is that their prompt emission light curves are probably periodically modulated with periods of a few to tens of seconds. 
\end{abstract}
\keywords{\uat{Hydrodynamical simulations}{767} --- \uat{Gamma-ray bursts}{629} --- \uat{X-ray transient sources}{1852} --- \uat{White dwarf stars}{1799} ---  \uat{Tidal disruption}{1696}}

\section{Introduction} \label{sec:intro}
$\gamma$-ray bursts (GRBs) can be classified into long and short GRBs based on their $T_{90}$ duration. It is generally believed that long GRBs ($T_{90}>2$ s) are associated with the core collapse of massive stars and accompanied by supernova explosions, while short GRBs ($T_{90}<2$ s) are linked to mergers of binary neutron stars (BNS) or NS-black hole (NS-BH) systems, typically accompanied by gravitational waves (GWs) and kilonovae \citep{2015PhR...561....1K, 2018pgrb.book.....Z}. Notably, the connection between BNS mergers and short GRBs has been observationally confirmed through GW detections \citep{2017PhRvL.119p1101A, 2017ApJ...848L..13A}.

Certain peculiar cases, such as GRB 230307A and 211211A, have generated substantial scientific intrigue. Their defining features include:
(1) $T_{90}>2$ s, often reaching tens of seconds \citep{2022Natur.612..223R, 2022Natur.612..228T, 2022Natur.612..232Y, 2024Natur.626..737L, 2024Natur.626..742Y, 2025NSRev..12E.401S, 2025ApJ...985..239Y};
(2) consistently low redshifts \citep{2022Natur.612..223R, 2022Natur.612..228T, 2024Natur.626..737L, 2024Natur.626..742Y};
(3) significant location offsets from their host galaxies (e.g., GRB 230307A shows an offset of $\sim$ 37 kpc, \citeauthor{2024Natur.626..737L} \citeyear{2024Natur.626..737L}, \citeauthor{2024Natur.626..742Y} \citeyear{2024Natur.626..742Y});
(4) absence of late-time supernova signatures despite exhibiting kilonova-like components similar to AT 2017gfo \citep{2022Natur.612..223R, 2024Natur.626..742Y};
(5) potential quasi-periodic modulation features in the prompt emission \citep{2024ApJ...973L..33C};
(6) Other potential observational features, such as the visible dips between prompt and extended emission, the nearly constant ratio of extended to main emission energy, and an ever decreasing hardness over time \citep{2025NSRev..12E.401S, 2024ApJ...969...26P, 2025ApJ...979...73W, 2025arXiv250406616T, 2025JHEAp..45..325Z, 2025A&A...698A.250K}, point to a perplexing origin for this type of long-merger GRBs.

BNS or NS-BH mergers are generally not expected to produce GRBs with durations of tens of seconds in standard accretion-powered scenarios. \cite{2025JHEAp..45..325Z} points out that the duration of a GRB may be regulated by multiple factors, including the progenitor system, the central engine, the emitter, and the jet geometry. Under this framework, BNS or NS-BH systems could, in principle, produce GRBs with longer durations under certain conditions.

As a potential progenitor, white dwarf (WD)-BH/NS mergers provide a new perspective for understanding the origin of long-merger GRBs  \citep{1999ApJ...520..650F, 2007MNRAS.374L..34K, 2018MNRAS.475L.101D, 2022Natur.612..232Y, 2023ApJ...947L..21Z, 2024MNRAS.535.2800L, 2024ApJ...964L...9W, 2025ApJ...988L..46L, 2025arXiv250810984C}. Moreover, compared with BNS or NS-BH systems, WD-BH/NS binary systems are more abundant in the universe. Therefore, their merger process merits in-depth investigation.

Using three-dimensional hydrodynamical simulations, \cite{1999ApJ...520..650F} studied the merger process of a WD with a BH or NS, modeling the WD with 6,000-16,000 particles and focusing specifically on its tidal disruption process. Their findings indicate that once the stripped mass of the WD exceeds $\sim 0.2\, M_{\odot}$, the remnant material drains into the accretion disk around the NS or BH within one orbital period, which results in a peak accretion rate of $\sim 0.05\, M_{\odot} \ \mathrm{s}^{-1}$, suggesting that it could potentially produce long GRBs. 

Since then, considerable efforts have been devoted to investigating the evolution of accretion disks formed in WD-BH/NS systems \citep{2012MNRAS.419..827M, 2016MNRAS.461.1154M, 2018MNRAS.475L.101D, 2019MNRAS.486.1805Z, 2020MNRAS.493.3956Z, 2022MNRAS.510..992T, 2022MNRAS.510.3758B, 2023ApJ...956...71K, 2024A&A...681A..41M}. Simulations by \cite{2024A&A...681A..41M} demonstrate that the tidal disruption of the WD by the NS leads to the formation of an accretion disk, and that subsequent accretion onto the NS powers a strongly magnetized, mildly relativistic jet launched perpendicular to the orbital plane. \cite{2019MNRAS.486.1805Z} investigated the disruption of CO-WDs by NSs, and found that the ejecta contained only a trace amount of $^{56}$Ni (few $\times 10^{-3} \,M_{\odot}$), giving rise to faint, rapidly evolving, and reddened transients. Such events are substantially shorter-lived and considerably fainter than both regular and faint/peculiar Type Ia supernovae. A recent work by \cite{2026A&A...706A.106K} shows that the intrinsically non-spherical wind ejecta produced in WD-NS mergers can lead to strong viewing-angle dependence in $^{56}\mathrm{Ni}$-powered thermal transients, with the same event exhibiting substantially different luminosities and temporal and spectral properties when observed from different directions.

The candidate kilonova features found in those long-merger type GRBs were generally thought to be associated with r-process nucleosynthesis of super-heavy elements. However, \cite{2025arXiv250903003R} found that an ejecta with a weak r-process component, though lacking lanthanide-rich material, can also reproduce the late-time optical and near-infrared (NIR) light curves observed for GRB 211211A and 230307A (see also \citeauthor{2025ApJ...982...81M} \citeyear{2025ApJ...982...81M}). Furthermore, varying assumptions about the velocity distribution of a lanthanide-lacking ejecta can also produce late-time features that mimic phenomena typically associated with heavy-element synthesis \citep{2023ApJ...958..121T, 2024ApJ...961....9F}. 

\defcitealias{2024ApJ...973L..33C}{Paper I} 
\citetalias{2024ApJ...973L..33C} (\citeauthor{2024ApJ...973L..33C} \citeyear{2024ApJ...973L..33C}; hereafter \citetalias{2024ApJ...973L..33C}) pointed out that, due to the relatively large tidal disruption radius of a WD, a non-negligible residual orbital eccentricity $e$ may be retained during the late stages of its merger with a NS or BH. Consequently, the WD can experience repeated partial disruptions (RPDs) by the BH or NS. They further found quasi-periodic modulations in the prompt emission of long-merger GRBs (such as GRB 230307A, 211211A, and 060614), supporting this scenario.

In this work, we aim to numerically study in detail the RPD scenario proposed in \citetalias{2024ApJ...973L..33C}. We use high-resolution SPH codes to simulate the late-stage evolution of WD mergers with BHs. We consider different initial WD masses and residual orbital eccentricities. Our focus encompasses the dynamical evolution of the binary system, the structural evolution of the WD, the mass transfer, and potential accretion. Note that although our simulations primarily focus on WD-BH systems, the process is also applicable to WD-NS systems, particularly for simulations in which the compact companion mass is set to $2\,M_{\odot}$. We also explore possible observational signatures arising from the WD-BH/NS merger, including X-ray transients and GRBs.

The paper is organized as follows. \S \ref{sec2} briefly reviews the physical scenario. \S \ref{sec3} describes the methodology. \S \ref{sec4} presents the simulation results. \S \ref{sec5} discusses the observational appearance, including the jet's generation, power, and radiation. Conclusions are provided in \S \ref{sec6}. 

\section{Physical Scenario} \label{sec2}

A WD-BH/NS system can form in two channels, in both of which significant orbital eccentricities may appear. In the isolated stellar binary evolution channel, if the primary has an initial mass $\gtrsim 8\, M_\odot$, during the late stage of stellar evolution, it eventually forms a NS or BH in a core-collapse explosion. A large-scale mass ejection from the explosion can significantly alter the binary's orbit; even a small amount of ejected material may potentially lead to a significant increase in the orbital eccentricity. 

WD-BH/NS systems can also form via dynamical capture, particularly in dense stellar environments such as globular clusters, where the resulting binaries naturally exhibit high orbital eccentricities.

\begin{figure}[htbp]
    \centering
    \includegraphics[height=6cm,width=8.5cm]{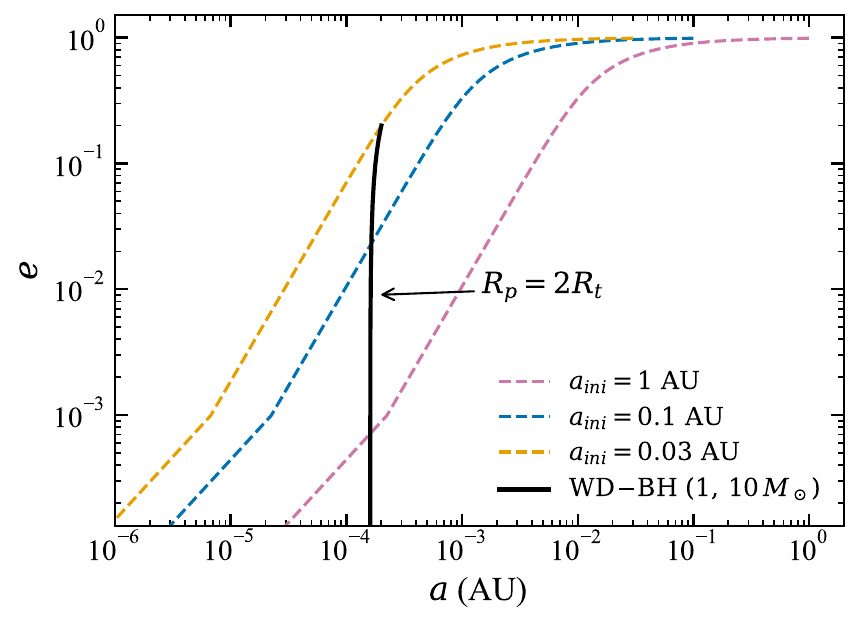}
    \caption{Evolution of the semi-major axis $a$ and eccentricity $e$ of  compact binaries driven by the GW radiation (dashed lines). The initial eccentricities are set to $e_{\rm ini} = 0.99$. Different colors represent different initial $a$'s. The black solid line represents the $a$-$e$ relation at the onset of mass transfer ($R_p \sim 2R_t$) for a WD-BH system, i.e., the ending point of its GW-driven orbital evolution. } 
    \label{Fig: e_a}
\end{figure}

Using formalism in \cite{1964PhRv..136.1224P} we calculate and show in Figure \ref{Fig: e_a} the GW-driven orbital evolution of a binary composed of stellar-mass compact objects, for given initial eccentricity of $e_{\rm ini} \simeq 0.99$ and the initial semi-major axis in the range $0.03 \leq  a_{\rm ini} \leq  1\,\mathrm{AU}$. For a NS-BH or BNS system, its orbit would be fully circularized by the time mass transfer begins (at $a\lesssim 10^{-6}$ AU). In contrast, owing to the relatively large tidal disruption radius of WDs, WD-BH/NS systems may retain a residual orbital eccentricity in the range $0 \lesssim e_0 \lesssim 0.2$ at the onset of mass transfer.

\subsection{Start of RPD} \label{sec2_0}

For a binary composed of a WD with mass $M_*$, and a BH or NS with mass $M$, in an orbit of semi-major axis $a$ and eccentricity $e$, the total angular momentum and total orbital energy of the binary are 
$J_{\rm orb}=M_* M /(M_*+M) \times \sqrt{G(M_*+M)R_p(1+e)}$  and $E_{\rm orb}=-G M_* M /(2a)$,
respectively, where $R_p = a(1-e)$ is the peri-center radius. 

The mass–radius relation of the WD is  \citep{1972ApJ...175..417N}
\begin{equation}
R_*=9\times10^8  \left(\frac{M_*}{M_{\odot}}\right)^{-1/3}\left[1-\left(\frac{M_*}{M_{\rm Ch}}\right)^{4/3}\right]^{1/2} \ {\rm cm},
\label{eq:R_*} \end{equation}
where $M_{\rm Ch} \simeq 1.44 M_\odot$. At the the peri-center, the instantaneous Roche lobe radius of the WD is $R_L= 0.49 \, q^{2/3} R_p/[0.6 \, q^{2/3}+\ln (1+q^{1/3})]$, where $q=M_*/M$ is the mass ratio \citep{2007ApJ...660.1624S}. 

When $R_*>R_L$, the WD instantaneously overflows its Roche lobe, causing part of its mass to be stripped by the compact companion star. For orbits retaining some eccentricity, Roche lobe overflow occurs predominantly near $R_p$.
Consequently, both mass transfer and subsequent accretion processes are modulated at the orbital period. This indicates that the WD undergoes RPD by the BH/NS. 

One can also define a disruption radius of the WD as \citep{1971ARA&A...9..183P, 1988Natur.333..523R}
\begin{equation}
R_t  =R_* \left(\frac{M}{M_*}\right)^{1/3},
\label{eq:R_t} \end{equation}
and introduce the penetration factor $\beta = R_t / R_p$ to quantify the degree of Roche-lobe overflow. The onset condition for the Roche-lobe overflow $R_* = R_L$ corresponds to a critical penetration $\beta_0 \approx 0.5$.

The orbital period at the RPD onset is given by Kepler's third law as
\begin{multline}  \label{eq:period}
P_0= 50 \, \left(\frac{\beta_0}{0.5}\right)^{-3/2} (1-e)^{-3/2} (1+q)^{-1/2} \\ \times \left(\frac{M_*}{M_\odot}\right)^{-1} \left(1- \frac{M_*}{M_{\rm Ch}}\right)^{0.67} \,\mbox{s} ,  
\end{multline} 
where we have used the approximate form of $R_* \propto M_*^{-1/3} (1- M_*/M_{\rm Ch})^{0.447}$ by \cite{2010MNRAS.409L..25Z}
to Eq.~(\ref{eq:R_*}). Note that this period depends primarily on $M_*$ only.

\subsection{Orbital evolution during RPD} \label{sec2_1}

In the absence of perturbation from a third body, the orbital evolution of a WD-BH/NS binary is governed primarily by GW radiation, tidal dissipation, and mass transfer. 


The GW-driven orbital shrinkage timescale is defined as $t_{\rm GW}(a) \equiv a/|\dot{a}| \equiv E/|\dot{E}|$, where $E$ is the orbital energy and $\dot{E}$ is the orbital mean of GW radiation power \citep{1983bhwd.book.....S, 1963PhRv..131..435P} 
\begin{equation}
\dot{E}=-\frac{32}{5} \frac{G^4}{c^5} \frac{(M_*+M)^3 M_*}{a^5} \frac{f(e)}{(1-e^2)^{7/2}},
\label{eq:dotE} \end{equation}
where $f(e)=1 +(73/24) e^2 +(37/96) e^4$. Then it can be estimated as
\begin{multline} \label{eq:tgw} 
t_{GW}\simeq 7.8 \ m_*^{-1} \left( \frac{P}{20 \ {\rm s}}\right)^{8/3} \left( \frac{M} {10 M_\odot} \right)^{-2/3} \\
\times \frac{(1-e^2)^{7/2}}{f(e)} \ {\rm yr}.
\end{multline}
Clearly, for a WD with an orbital period of $\sim 20\,\mathrm{s}$, it would still take a few years for the WD to merge with a BH or a NS if via GW radiation alone. Therefore, the effect of GW emission on the orbital evolution during the RPD phase is extremely weak and can be safely neglected.

Tidal dissipation also reduces orbital energy. Tidal forces on the surface of the WD induce non-uniform stretching and compression, thereby exciting internal non-radial oscillation modes and converting part of the orbital energy into the WD's oscillation energy \citep{1977ApJ...213..183P}. The injected energy is approximately proportional to the square of the oscillation amplitude \citep{2011ApJ...732...74G}, $\displaystyle| \delta E_*/E_{*} |\approx ( \delta R /R)^2$, where $E_* \approx G M_*^2/R_*$ is the gravitational binding energy of the star. 

For mass transfer, under the approximation of angular momentum conservation (i.e., $\dot{J} \sim 0$ and $\dot{M}_* \sim -\dot{M}$), material flows from the WD to the BH or NS, while angular momentum is transferred back to the WD, resulting in an increase of the $R_p$ and an expansion of the Roche lobe $R_{\rm Lobe}$. If the WD expands faster than its Roche lobe, $\Delta R_* > \Delta R_L$, the mass transfer becomes unstable; otherwise, it remains stable \citep{2017MNRAS.467.3556B, 2017ApJ...851L...4C, 2018MNRAS.475L.101D, 2024MNRAS.529.1347D}. 

In a typical run of our simulations, e.g., the merger of a $1\,M_\odot$ WD with a $10\,M_\odot$ BH with $\beta = 0.5$, one finds that for tidal dissipation, $\delta E_*/E_{*} \sim 10^{-4}$, while transferring a mass of  $0.1\,M_\odot$ would cause $\Delta {a}/{a} \sim 0.2$. Therefore, during the RPD phase, mass transfer almost dominates entirely the orbital evolution.

\begin{deluxetable*}{cccccccccccccc} 
\tabletypesize{\scriptsize}
\tablewidth{0pt} 
\tablecaption{Initial parameters and results of our simulation runs. \label{tab:tab1}}
\tablehead{
\multicolumn{1}{c}{\textbf{Run ID}} &
\multicolumn{7}{c}{\textbf{Initial Parameters}\tablenotemark{a}} & 
\multicolumn{5}{c}{\textbf{Simulation Results}\tablenotemark{b}} \\
\cline{1-1} \cline{2-8} \cline{9-14}
\noalign{\vskip 1pt}
\colhead{No.} & 
\colhead{$M_*$}         & \colhead{$M$}    & \colhead{$q$}         & \colhead{$\beta_0$}  & \colhead{$e_0$} & \colhead{$a_0$}         & \colhead{$P_0$} & \colhead{RPD Duration}  & \colhead{$\dot{M}_{\rm mean}$} & \colhead{$\dot{M}_{\rm peak}$}  & \colhead{$\Delta M$} &  \colhead{$\Delta a_\bullet$} &  \colhead{Category}\\
 & ($M_{\odot}$) & ($M_{\odot}$) &  & 
 &  & ($R_{\odot}$) & (s) & (s) &
($M_{\odot}$/s) & ($M_{\odot}$/s) & ($M_{\odot}$) &  & 
}
\startdata
1a  & 0.8   & 10 & 0.08 & 0.45 & 0.2  & 6.6E-2 & 51.9 & 1190 ($\sim$ 21 $P$)  & 2.1E-4 &8.1E-4 & 0.25 & 0.08 &  III  \\
2a  & 1      & 10 & 0.1 & 0.45 & 0.2  & 4.8E-2 & 31.2 & 615 ($\sim$ 18 $P$)  & 6.3E-4 & 3.4E-3 & 0.40 & 0.13 & \\
4  & 1      & 10 & 0.1 & 0.4 & 0.2  & 5.4E-2 & 37.9 & 2800 ($\sim$ 65 $P$)  & 8.9E-5 & 3.8E-4 & 0.23 & 0.08 & \\
6a  & 1      & 10 & 0.1 & 0.45 & 0.05 & 4.0E-2 & 24.6 & 825 ($\sim$ 26 $P$)  & 5.0E-4 &  1.3E-3 & 0.42 & 0.14 & \\
7a  & 1      & 10 & 0.1 & 0.45 & 0.1  & 4.3E-2 & 26.6 & 1040 ($\sim$ 31 $P$)  & 3.9E-4 & 1.4E-3 & 0.41 & 0.14  & \\
\hline
8a  & 0.8   & 2  & 0.4 & 0.45 & 0.2  & 3.9E-2 & 45.6 & 420 ($\sim$ 9 $P$) & 1.1E-3 & 3.9E-3  & 0.40 & 0.58 (0.31) &  II \\
0a  & 1   & 10 & 0.1 & 0.45 & 0.01  & 3.8E-2 & 22.7 & 220 ($\sim$ 9 $P$)  & 1.2E-3 & 2.2E-3 & 0.28 & 0.09 & \\
2  & 1      & 10 & 0.1 & 0.5 & 0.2  & 4.3E-2 & 27.1 & 99 ($\sim$ 3 $P$) & 1.6E-3 & 6.0E-3  & 0.17 & 0.06 & \\
3a  & 1.3   & 10 & 0.13 & 0.45 & 0.2  & 2.3E-2 & 10.5  & 270 ($\sim$ 23 $P$)& 2.0E-3 & 6.5E-3  & 0.57 & 0.17 & \\
5  & 1      & 10 & 0.1 & 0.6 & 0.2  & 3.6E-2 & 20.6 & 38 ($\sim$ 2 $P$) & 3.6E-3 &7.5E-3  & 0.15 & 0.05 & \\
6  & 1      & 10 & 0.1 & 0.5 & 0.05 & 3.6E-2 & 21.0 & 43 ($\sim$ 2 $P$)  & 4.9E-3 & 1.2E-2 & 0.22 & 0.07 & \\
7  & 1      & 10 & 0.1 & 0.5 & 0.1  & 3.8E-2 & 22.7 & 62 ($\sim$ 3 $P$) & 2.5E-3 & 9.0E-3  & 0.17 & 0.06 & \\
\hline
10a & 1.3   & 2  & 0.65 & 0.45 & 0.2  & 1.4E-2 & 8.7 & 27 ($\sim$ 3 $P$)& 2.2E-2 & 8.5E-2  & 0.66 & 0.86 (0.43) & I \\
9  & 1      & 2  & 0.5 & 0.5 & 0.2  & 2.5E-2 & 23.2 & 43 ($\sim$ 2 $P$) & 1.0E-2 & 2.4E-2 & 0.43 & 0.61 (0.32) & \\
10 & 1.3   & 2  & 0.65 & 0.5 & 0.2  & 1.2E-2 & 7.4  & 12.2 ($\sim$ 1 $P$) & 4.8E-2 &  0.13& 0.60 & 0.8 (0.41) & \\
11 & 1.4     & 5  & 0.28 & 0.5 & 0.2  & 8.5E-3 & 3.1 & 7.5 ($\sim$ 2 $P$)  & 5.3E-2 & 0.16& 0.42 & 0.27 & \\
\enddata  
\vspace{-4pt}
\tablenotetext{a}{$M_*$, $M$ and $q$ are the WD mass, the BH or NS mass and the mass ratio, respectively. $\beta_0$, $e_0$, $a_0$ and $P_0$ are the initial orbital penetration parameter, eccentricity, semi-major axis and period, respectively, at the start of the simulation.} 
\vspace{-4pt}
\tablenotetext{b}{RPD duration is defined as the time interval from the start of the simulation to the final disruption of the WD. $\dot{M}_{\rm mean}$, $\dot{M}_{\rm peak}$ and $\Delta M$ are the mean and peak accretion rates, and the total mass accreted during the RPD phase, respectively. $\Delta a_{\bullet}$ denotes the total increase of the dimensionless spin parameter due to mass accretion. In the BH case, it is estimated by multiplying the accreted mass by the specific Keplerian angular momentum at the innermost stable circular orbit. Values in parentheses correspond to estimates using the specific Keplerian angular momentum at the NS surface when the accreting object is a NS. The simulation runs are classified into three categories according to the values of $\dot{M}_{\rm mean}$.}
\end{deluxetable*}

\section{Methods} \label{sec3}

We use the SPH code Phantom (\citeauthor{2012JCoPh.231..759P} \citeyear{2012JCoPh.231..759P}; \citeauthor{2018PASA...35...31P} \citeyear{2018PASA...35...31P}, for a detailed description of Phantom and the methodology, see \citeauthor{2019MNRAS.485..819L} \citeyear{2019MNRAS.485..819L}). The initial WD is modeled as a polytrope with index $n=3/2$, with its mass–radius relation determined by Eq. (\ref{eq:R_*}). In all simulations, the WD is discretized into $10^6$ equal-mass particles, while the NS or BH is represented by a sink particle, acting as a point mass for gravitational interactions. 

Throughout the simulation, both the gas in the stripped stream and the remaining WD are assumed to follow a polytropic equation of state (EOS) with an index $\gamma = 5/3$ \citep{2025MNRAS.542..839G}.  This value of $\gamma$ is consistent with the effective equation of state of a non-relativistic degenerate electron gas, which provides the dominant pressure support in a typical WD. 

The WD is placed on an elliptical orbit, initially near the aphelion $R_a$, with a velocity $v_0 = \sqrt {2GM(1-e)/R_a}$, and its orbital motion is counterclockwise in the $x$-$y$ plane. Since the WD remains at relatively large distances from the BH during mass transfer ($R_p \geq 100\,R_s$), general relativistic effects are neglected. To avoid extremely small time steps near the BH event horizon, particles that fall into a fixed accretion radius (set to $R_{\rm acc} = R_p/3$, see \S \ref{sec4_3}) are immediately removed.

\subsection{Initial parameters} \label{sec3_1}

WD-NS or WD-BH binaries gradually lose orbital energy and angular momentum through the GW emission over the timescale $t_{\rm GW}$. In our simulations, we adopt two approximate values of $\beta_0$, 0.45 and 0.5, as the initial conditions. We also had two additional runs with $\beta_0 = 0.4$ and $0.6$, respectively, to investigate our results' dependence on $\beta_0$ (see \S \ref{sec4_4}).

All simulations focus exclusively on massive WDs, with their initial masses set to $M_*=$ 0.8, 1.0, and 1.3 $M_{\odot}$, respectively. The BH mass is fixed at $M=$ 10 $M_{\odot}$ and 5 $M_{\odot}$, respectively, while NSs are approximated as point masses of $M= 2 \, M_{\odot}$. Initial orbital eccentricities of $e_0 =$ 0.05, 0.1, and 0.2, respectively, are adopted based on the estimate in \S \ref{sec2}. For demonstrating the role of eccentricity (see \S \ref{sec4_4}), we had an additional run with $e_0 =0.01$. 

Table \ref{tab:tab1} lists all 16 simulation runs, along with their initial parameters and simulation results.

\section{result} \label{sec4}

We start with the run of the merger of a 1 $M_{\odot}$ WD and a 10 $M_{\odot}$ BH as a representative case, presenting the simulation results on the structure of the WD (\S \ref{sec4_1}) and the orbital evolution (\S \ref{sec4_2}). \S \ref{sec4_3} and \ref{sec4_4} respectively present the mass accretion processes from several simulations and analyze the parameter dependence of the results within the simulation ensemble.

\subsection{WD structure} \label{sec4_1}
\begin{figure*}[htbp]
    \centering
    \includegraphics[height=6cm,width=14.0cm]{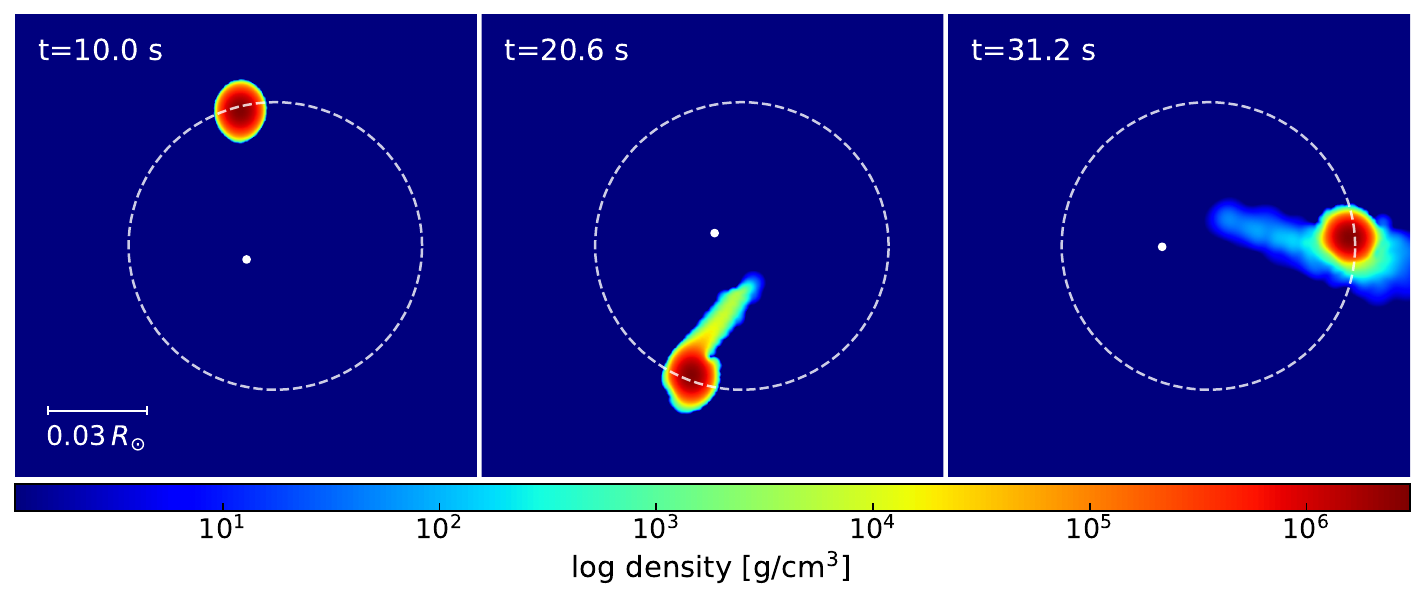}
    \caption{Density evolution of the WD over one orbital period. We present density‐projection snapshots in the $x$–$y$ plane from a simulation using $10^{6}$ SPH particles to model the late merger phase of a $1\,M_{\odot}$ WD interacting with a $10\,M_{\odot}$ BH (for simulation runs No. 2a). Each panel covers a region of \(0.14 \times 0.14\,R_{\odot}\).}
    \label{Fig: binary_orbit_one}
\end{figure*}
The binary system in the simulation run No. 2a (with $\beta_0$= 0.45, $e_0$ = 0.2, $P_0$ = 31.2 s, see Table \ref{tab:tab1}), exhibits distinct evolutionary phases during its first orbital period as captured by time-equidistant snapshots on the $x$-$y$ plane shown in Figure~\ref{Fig: binary_orbit_one}. At simulation onset ($t=$  0 s), the WD occupies $R_a$, its trajectory marked by the white dashed line. Incipient tidal deformation emerges at $t=$ 10.0 s, preceding a morphological transition at $R_p$ passage ($t=$ 20.6 s) where the WD evolves from spherical to an ellipsoidal shape with pronounced teardrop distortion, simultaneously developing a substantial debris stream directed toward the BH. 

Subsequent evolution at $t=$ 31.2 s (right panel of Figure~\ref{Fig: binary_orbit_one}) reveals structural reconvergence toward sphericity for the WD at $R_a$, while stripped material bifurcates: one component accelerates toward the BH for accretion, whereas the remaining debris is gravitationally pulled to the WD. 

Determining the exact amount of residual material that remains bound to the WD after experiencing a partial disruption is a highly challenging task. As an alternative approach, \cite{2025ApJ...979...40L} defined a threshold surface density ($\rho_t$) established after relaxation, where all the material with $\rho > \rho_t$ is considered bound to the star. Adopting this approach, we determine the residual structure of the WD after each stripping and hence the new $M_*$ and $R_*$. 

\subsection{Orbital evolution} \label{sec4_2}
\begin{figure*}[htbp]
    \centering
    \begin{interactive}{animation}{2a_RPD.mp4}
    \includegraphics[height=12.0cm,width=12.0cm]{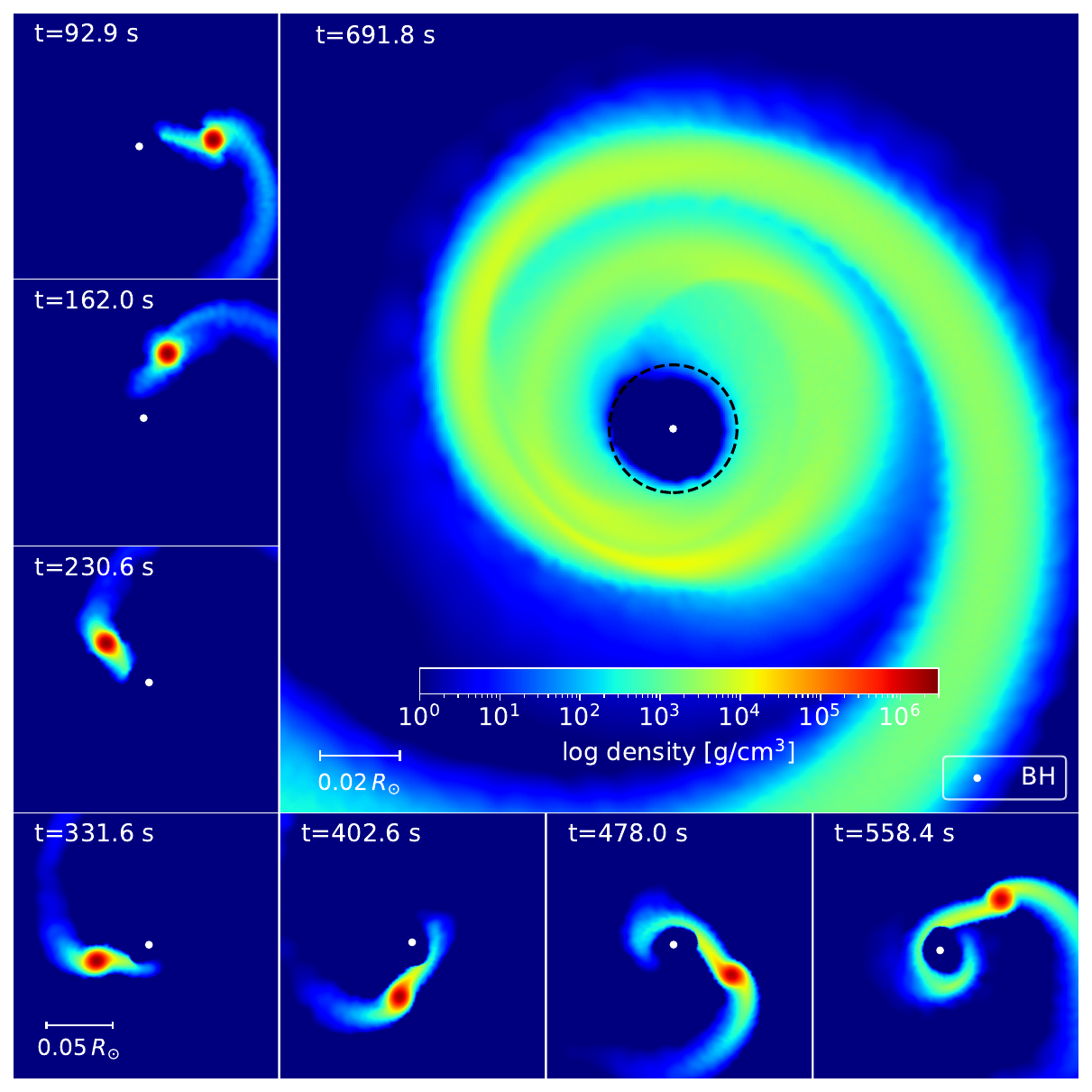}
    \end{interactive}
    \caption{Repeating Partial Disruption Process of a WD. We present density‐projection snapshots in the $x$-$y$ plane from a simulation using $10^{6}$ SPH particles to model the late merger phase of a $1\,M_{\odot}$ WD interacting with a $10\,M_{\odot}$ BH (simulation run No. 2a). Each panel corresponds to a different evolutionary time, with all panels spanning a physical scale of $0.2\times0.2\,R_{\odot}$. The main panel shows the WD after complete disruption (the termination of the RPD is defined as $t=620$ s), where the black dashed circle marks the accretion radius $R_{\rm acc}= R_p/3$. \\ (An animation of this figure is available in the online article.)} \label{Fig: binary_evolution}
\end{figure*}

Figure \ref{Fig: binary_evolution} presents a series of density snapshots in the $x$–$y$ plane at representative orbital phases during the No. 2a run. Two distinct debris streams are observed to extend from the two sides of the WD, with the stream directed toward the BH being gravitationally captured by the latter. Over successive orbits, spiral arms develop and exhibit progressive density enhancement on both flanks of the WD, thereby intensifying the partial disruption process. This evolutionary pattern is consistent with the scenario predicted by Figure 1 of \citetalias{2024ApJ...973L..33C}.

\begin{figure}[htbp]
    \centering
    \includegraphics[height=9.2cm,width=8.5cm]{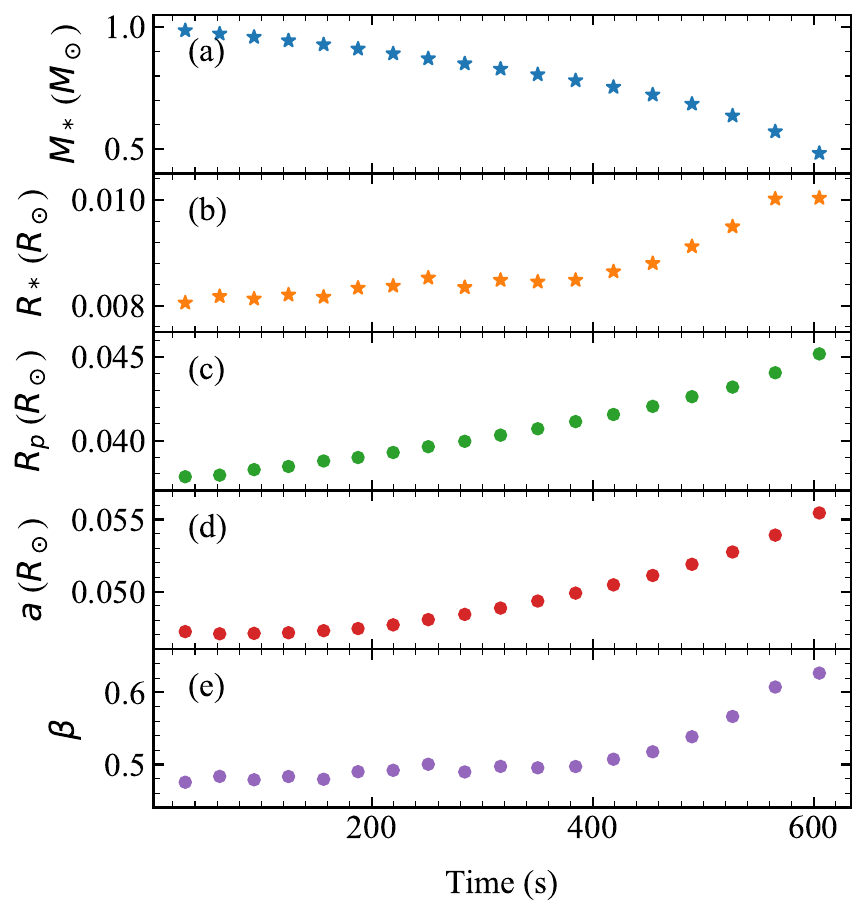}
    \caption{Late-stage parameter evolution of a 1 $M_{\odot}$ WD merging with a 10 $M_{\odot}$ BH (for simulation runs No. 2a). Panels (a), (b), (c), (d), and (e) show the evolution of WD’s mass $M_*$, radius $R_*$, pericenter distance $R_p$, semi-major axis $a$, and penetration factor $\beta$, respectively.}
    \label{Fig: merger_wd_evolution}
\end{figure}

Panels (a)-(e) of Figure~\ref{Fig: merger_wd_evolution} show the evolution of $M_*$, $R_*$, $R_p$, $a$, and $\beta$ during the RPDs ($t = 0-620$ s, for the No. 2a run). As the WD undergoes RPDs, its mass $M_*$ gradually decreases, while the WD expands and its radius $R_*$ increases. The orbital parameters, $R_p$ and $a$, also grow with time, but $\beta$ exhibits an overall increasing trend. This behavior implies that the RPDs proceed through an unstable mass-transfer process, ultimately leading to the complete disruption of the WD. 

At $t \approx 620$ s (at $R_a$), the undergoes complete structural disruption, beyond which it can no longer maintain hydrostatic equilibrium; we define this as the termination of the RPD. The main panel of Figure~\ref{Fig: binary_evolution} shows the density distribution at $t = 691.8$ s, when the WD structure has almost entirely disappeared and the remaining debris is streaming toward the BH. A disk-like structure forms around the BH, and matter is continuously accreted into the accretion radius indicated by the black dashed circle in Figure~\ref{Fig: binary_evolution}.

\subsection{Mass Accretion History} \label{sec4_3}

The exact and detailed evolution of the accretion disk involves processes such as viscosity, radial drift of the disk material, and potentially neutrino cooling and nuclear burning \citep{2012MNRAS.419..827M, 2013ApJ...763..108F, 2016MNRAS.461.1154M} and is beyond the scope of this study. To simplify the accretion rate estimation, we adopt a relatively large accretion radius, $R_{\mathrm{acc}} \sim R_p/3$, and consider every particle that falls into $R_{acc}$ as being immediately accreted. 

\begin{figure}[htbp]
    \centering
    \includegraphics[height=6cm,width=8.5cm]{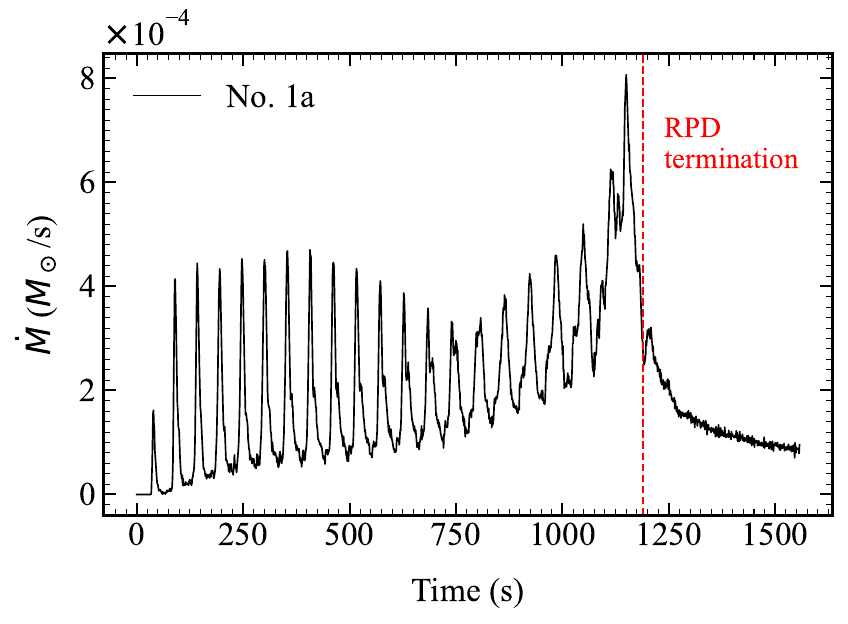}
    \includegraphics[height=6cm,width=8.5cm]{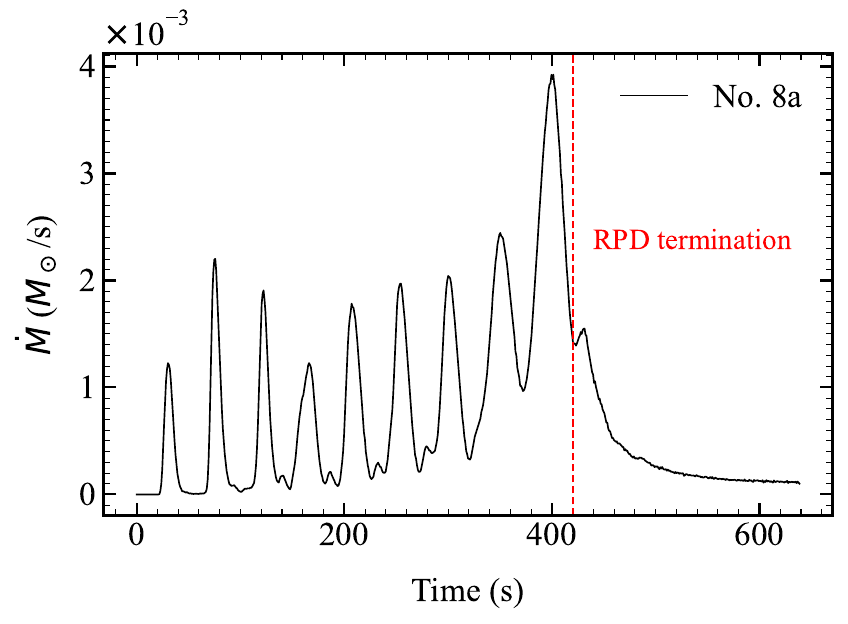}
    \centering
    \includegraphics[height=6cm,width=8.5cm]{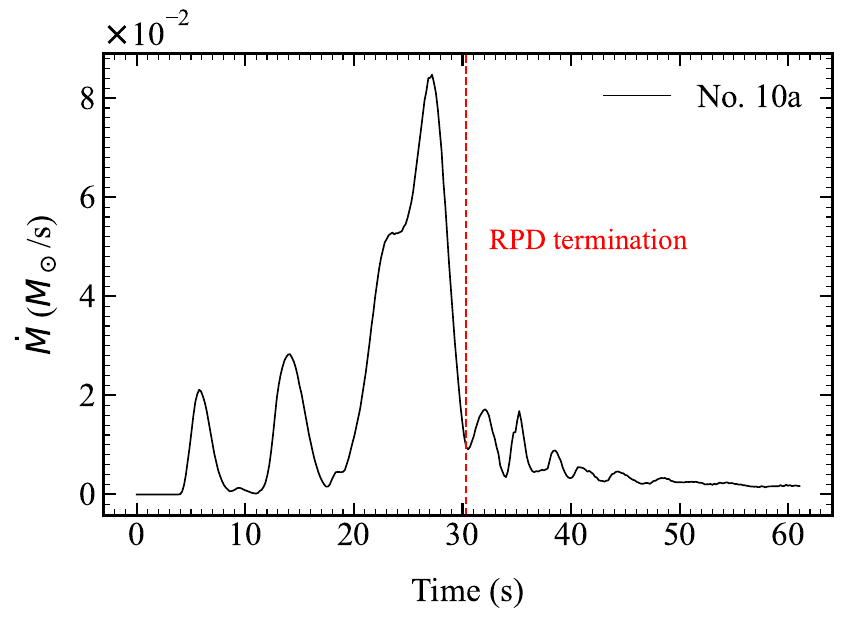}
    \caption{Evolution of the accretion rate during the late stage of the WD-BH merger for simulation runs No. 1a, 8a, and 10a. The red vertical dashed line indicates the termination of the repeated partial disruptions of the WD. Each panel corresponds to one of the three categories described in \S \ref{sec4_4}.}
    \label{Fig: accretion_rate}
\end{figure}
Figure \ref{Fig: accretion_rate} shows the evolution of the accretion rate $\dot{M}$ for simulations No. 1a, 8a, and 10a, respectively.
In simulation No. 1a (initial parameters $M_* = 0.8\, M_{\odot}$, $M = 10\,M_{\odot}$, $\beta_0 = 0.45$, and $e_0 = 0.2$), throughout the RPD phase, the accretion rate displays strong fluctuations on the orbital timescale. The peak accretion rate $\dot{M}_{peak}$ reaches $\sim 8.1 \times10^{-4}\, M_{\odot} \rm \ s^{-1}$ at $t \sim 1150$ s, when the WD structure is nearly destroyed. The RPD phase extends to about $t \sim 1200$ s ($\sim 21\,P$, indicated by the red vertical dashed line in Figure~\ref{Fig: accretion_rate}), after which the remnant material is accreted over a few orbital periods, with the $\dot{M}_{peak}$ gradually decreasing. 

For simulation No. 8a (initial $M_* = 0.8\, M_{\odot}$, $M = 2\, M_{\odot}$), compared with simulation No.1a, the larger $q$ leads to a markedly shorter RPD phase. At $t \sim 400$ s, the peak accretion rate reaches its $\dot{M}_{peak} \simeq$ $3.9 \times10^{-3} \,M_{\odot} \rm \ s^{-1}$. By $t \sim 420$ s ($\sim 9\,P$), the WD structure is nearly destroyed, marking the end of the RPD phase, followed by the subsequent accretion of the remaining debris material.

For simulation No. 10a (initial $M_* = 1.3\, M_{\odot}$, $M = 2 \,M_{\odot}$), the nearly-equal component masses and, most importantly, the much more compact WD lead to a much shorter orbital period compared with No. 1a and No. 8a. It has $\dot{M}_{peak} \simeq$ $8.5 \times 10^{2}\, M_{\odot} \rm \ s^{-1}$ at $t \sim 27$ s, and the RPD phase is highly violent, lasting for about $30$ s ($\sim 3\,P$).

We find that for all our simulations, the accretion history is strongly modulated by the orbital period, as is shown by the representative cases in Figure~\ref{Fig: accretion_rate}.

\begin{figure}[htbp]
    \centering
    \includegraphics[height=6cm,width=8.5cm]{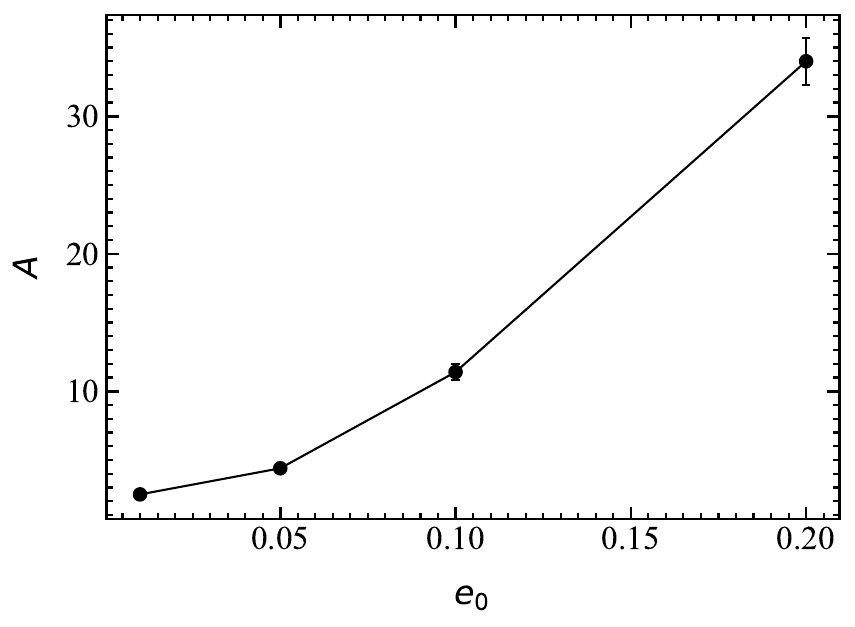}
    \caption{The relationship between the initial orbital eccentricity $e_0$ and the modulation amplitude $A = \dot{M}_{max}/\dot{M}_{min}$ obtained from simulation runs No. 0a, 2a, 6a, and 7a. The error bars represent the statistical standard errors of our measurements.}
    \label{Fig: modulation_index}
\end{figure}

The initial eccentricity $e_0$ plays a crucial role in modulating the accretion rate, as is shown in Figure~\ref{Fig: modulation_index}. We define the modulation amplitude as $A = \dot{M}_{max}/\dot{M}_{min}$, where $\dot{M}_{max}$ is the peak accretion rate and $\dot{M}_{min}$ is the minimum accretion rate within the same orbital cycle that contains the peak. Larger values of $A$ indicate stronger modulation. Figure \ref{Fig: modulation_index} depicts a clearly positive relation between $e_0$ and $A$ obtained from our simulations. 

However, the viscous evolution of the disk, not included in the simulations, may act to smooth out this modulation. The viscous timescale of the disk can be estimated as $t_{\rm vis} = R_p^2/\nu \simeq (P/ 2 \pi \alpha)(H/R)^{-2}$, where $\alpha$ is the Shakura–Sunyaev viscosity parameter and $H/R$ is the disk aspect ratio.
During the RPD phase, the accretion rate is extremely high, $\dot{M} \gg \dot{M}_{\rm Edd}$, so $H/R \simeq 1$, the disk is geometrically thick. Then for $\alpha \sim 0.1$, one would have $t_{vis} \approx P$. Therefore, we expect that the orbital modulation of the accretion rate may still leave imprints on the light curve.

\subsection{Classification and parameter dependence} \label{sec4_4}

Table \ref{tab:tab1} summarizes the peak ($\dot{M}_{peak}$) and mean ($\dot{M}_{mean}$) accretion rates, and RPD durations for all simulations. It shows a range of $10^{-4} \lesssim \dot{M}_{mean} \lesssim 10^{-1}\, M_{\odot} \rm \ s^{-1}$, and the RPD durations vary from 10 s to a few $\times 10^3$ s. 

In the decreasing order of their values of $\dot{M}_{mean}$, we classify all the simulation runs into three categories:
\begin{itemize}
    \item \textbf{Category I:} represented by run No. 10a (see Figure~\ref{Fig: accretion_rate}), including No. 9, 10, and 11, whose $\dot{M}_{mean}$ is $\gtrsim 10^{-2}\, M_{\odot} \rm \ s^{-1}$, with the RPD duration $\lesssim10^{2}$ s. 
    \item \textbf{Category II:} represented by run No. 8a, including No. 0a, 2, 3a, 5, 6, and 7, whose $\dot{M}_{mean}$ is intermediate: $10^{-3} \lesssim \dot{M}_{mean} \lesssim 10^{-2}\, M_{\odot} \rm \ s^{-1}$. The RPD phase typically lasts for $10^{2}$-$10^{3}$ s.
    \item \textbf{Category III:} represented by run No. 1a, including No. 2a, 4, 6a, and 7a. They are characterized by relatively low accretion rates, $\dot{M}_{mean} \lesssim10^{-3} \,M_{\odot} \rm \ s^{-1}$. The RPD phase can last more than $10^{3}$ s.
\end{itemize}

Compared with Categories II and III, Category I shows a larger modulation amplitude $A$, but with fewer RPD cycles.

\begin{figure}[htbp]
    \centering
    \includegraphics[height=6cm,width=8.5cm]{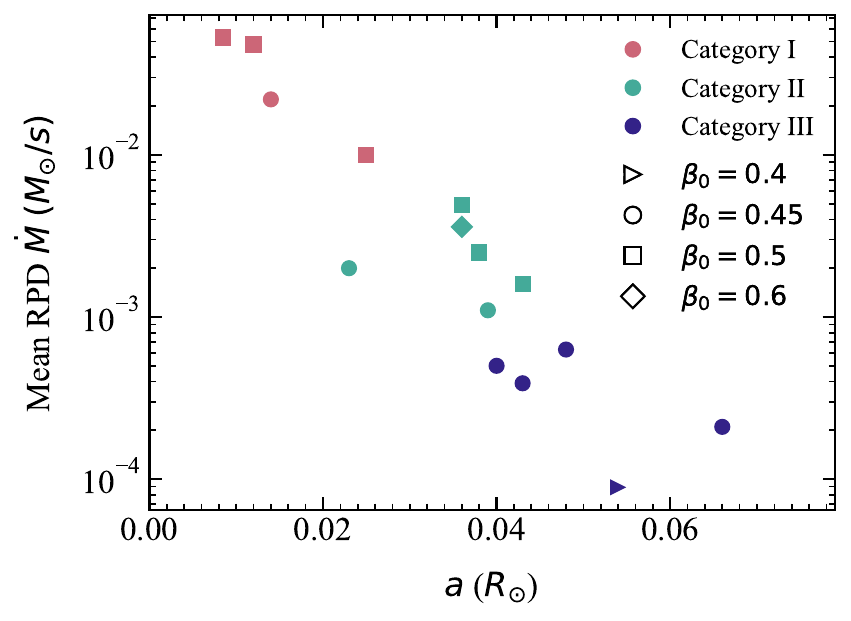}
    \includegraphics[height=6cm,width=8.5cm]{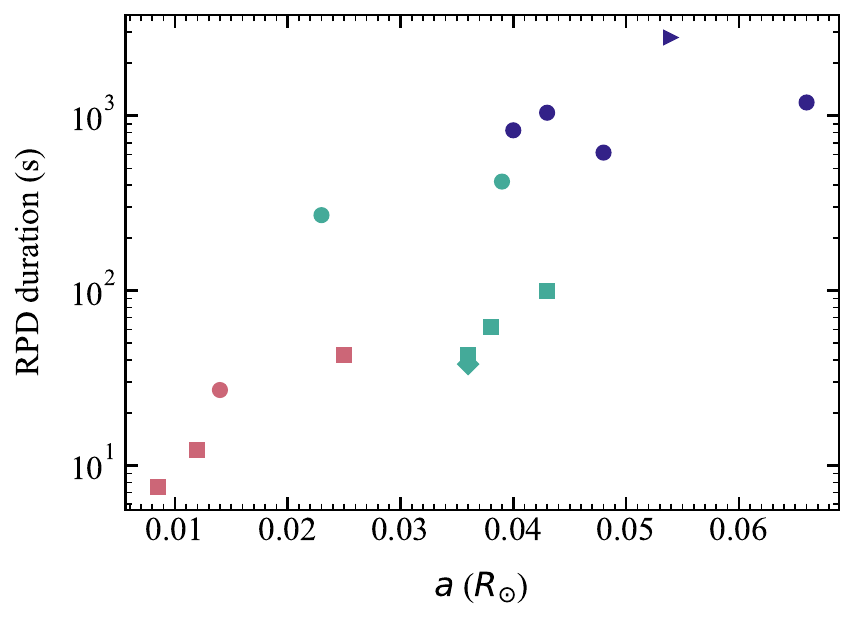}
    \centering
    \includegraphics[height=6cm,width=8.5cm]{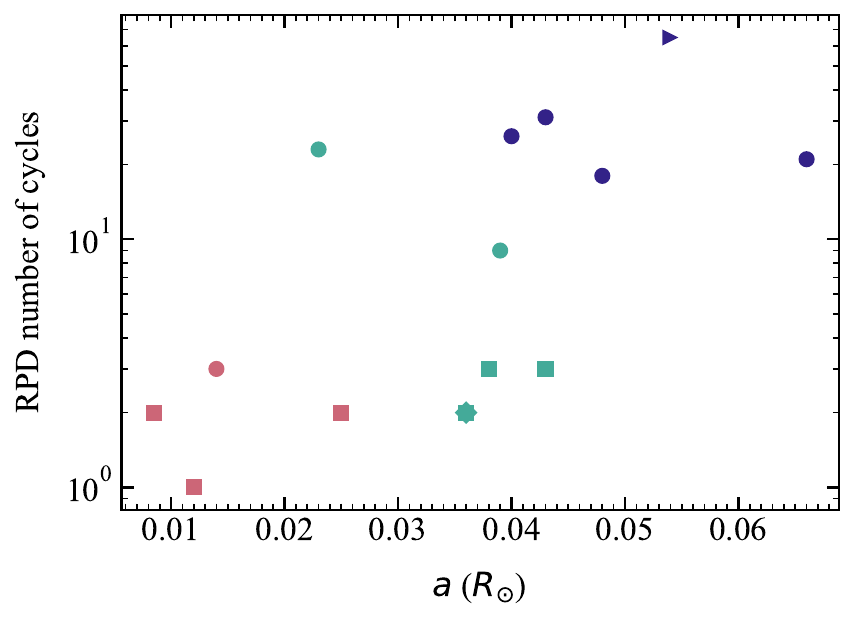}
    \caption{The relations of $\dot{M}_{mean}$, the RPD duration, and the number of RPD cycles versus the initial orbital semi-major axis $a$, among all our simulation runs. Red, green, and blue represent Categories I, II, and III, respectively. Triangles, circles, squares, and diamonds correspond to $\beta_0 = 0.4$, $0.45$, $0.5$, and $0.6$, respectively.} 
    \label{Fig: simulation_statistics}
\end{figure}

To delineate the factors behind the outcome variations among these categories, we plot in Figure~\ref{Fig: simulation_statistics} the relation of $\dot{M}_{mean}$, the RPD duration, and the number of RPD cycles versus the initial orbital size $a$. The main conclusions are: (1) $\dot{M}_{mean}$ decreases with increasing $a$; and (2) both the RPD duration and the number of cycles increase with increasing $a$. We also find similar trends of these three quantities of simulation outcomes with decreasing $q$ or increasing $P_0$, but they are somewhat weaker than those shown in Figure~\ref{Fig: simulation_statistics}. 

\begin{figure}[htbp]
    \centering
    \includegraphics[height=6cm,width=8.5cm]{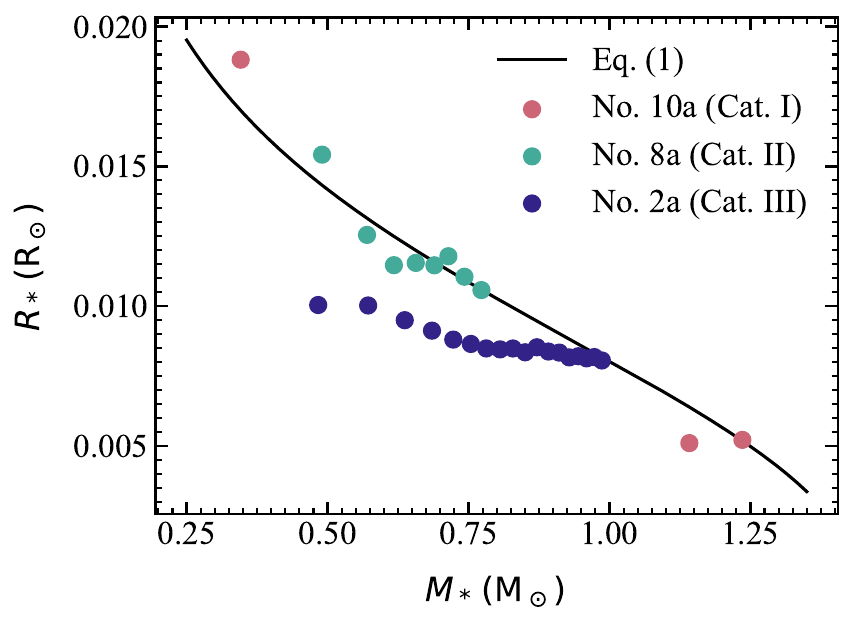}
    \caption{The successive evolution of the remnant WD mass $M_*$ and radius $R_*$ during the RPD process is shown for three representative simulation runs: No. 10a (Category I), 8a (Category II), and 2a (Category III). Their initial $M_*$ values are 1.3, 0.8, and 1.0 $M_\odot$, respectively. Following each partial disruption, the WD loses mass and undergoes expansion. The black line represents Eq.~(\ref{eq:R_*}), i.e., the mass–radius relation of an isolated, cold WD.} 
    \label{Fig: m_r}
\end{figure}

Figure \ref{Fig: m_r} plots the successive evolution of the remnant WD's radius and mass throughout the RPD process for simulations No. 10a, 8a, and 2a, corresponding to Categories I, II, and III, respectively. During the RPDs, as is shown in Figure~\ref{Fig: m_r}, the WD radius expansion rate $|dR_*/dM_*|$, which signifies the degree of unstableness of the mass transfer, somehow follows the sequence of Category I $>$ II $>$ III. This order explains why more compact systems -- with smaller $a$ or higher mass ratio $q$ -- exhibit fewer cycles, thus shorter durations and higher accretion rates. 

\begin{figure}[htbp]
    \centering
    \includegraphics[height=6cm,width=8.5cm]{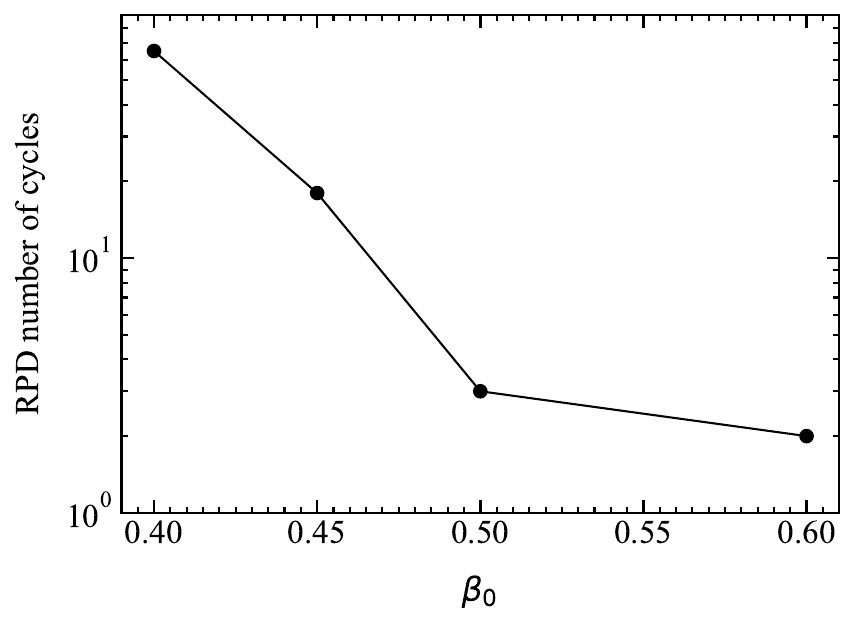}
    \caption{The relationship between the initial penetration factor $\beta_0$ and the number of RPD cycles obtained from simulation runs No. 2, 2a, 4, and 5 whose initial parameters are the same except for $\beta_0$.}
    \label{Fig: dura_beta0}
\end{figure}

We also plot in Figure~\ref{Fig: dura_beta0} the relation of the number of RPD cycles versus the initial penetration factor $\beta_0$, from simulation runs No. 2, 2a, 4, and 5. It is evident that the number of RPD cycles decreases with increasing $\beta_0$, indicating that the RPD process becomes more violent for deeper encounters.

\section{Observational Appearance } \label{sec5}

Based on our simulation results, here we discuss the observational consequences of a jetted WD-BH/NS merger.

In the RPD scenario of a WD-BH/NS merger, the central engine is an accreting BH or accreting NS. A jet could be launched, the mechanism of which is either neutrino-driven or magnetically driven, or both. We first review these in general (\S \ref{sec5_1}), paying attention to the jet luminosity's dependence on some parameters (accretion rate, spin, etc.). Then we estimate the properties of the jets in the context of our simulation results (\S \ref{sec5_2}), and make predictions about the jet emissions (\S \ref{sec5_3}). Finally we make connections to a few observed peculiar transients (\S \ref{sec5_5}).

\subsection{Central Engines and Jet Production} \label{sec5_1}

\subsubsection{BH engine}

When the central engine is a BH, the innermost region of the disk is so hot that copious neutrinos are produced there. Neutrinos ($\nu$) and antineutrinos ($\bar{\nu}$) can annihilate to produce photons and electron-positron pairs, while neutrinos may also strip baryons off the disk \citep{1996ApJ...471..331Q, 1999ApJ...518..356P, 1999ApJ...520..650F, 2017NewAR..79....1L}. A hot ``fireball" with modest baryon contamination is then formed above the accretion disk. Because the energy deposition region is located near the disk symmetry axis, where baryon loading is relatively low, the fireball can be efficiently accelerated to form a relativistic jet. The disk outflow or wind may further collimate the jet. 



Define $\dot{m}=\dot{M}/(M_\odot\,{\rm s^{-1}})$ as the dimensionless accretion rate. One usually adopts a critical accretion rate $\dot{m}_{\rm ign}=0.07-0.063a_\bullet$, where $a_\bullet = J_{\rm BH}c/(GM^2)$ is the dimensionless BH spin; below this ``ignition'' rate, the disk temperature would be too low to trigger neutrino-emitting reactions. Then the luminosity of a neutrino-annihilation driven jet is estimated as
\begin{multline}   \label{eq:Lvv}
    L_{\nu \bar{\nu},\, \rm BH} \simeq  
    10^{(48.0+0.15 a_\bullet)}
\left(\frac{m}{3} \right)^{\log\left(\frac{\dot{m}}{\dot{m}_{\rm ign}}\right) -3.3} \\
 \times \left[ \left(\frac{\dot{m}}{\dot{m}_{\rm ign}} \right)^{-4.7 }
+ \left(\frac{\dot{m}}{\dot{m}_{\rm ign}} \right)^{-2.23} \right]^{-1} \,{\rm erg\ s^{-1}},
\end{multline}
where $m = M/M_\odot$ \citep{2017ApJ...849...47L}. Note that the neutrino trapping by the disk is irrelevant here, since the peak accretion rates in all of our simulations are much lower than the corresponding trapping accretion rate \citep{2017ApJ...849...47L}.

On the other hand, if the accretion disk is highly magnetized and the rapidly spinning BH is threaded by magnetic field lines that connect the event horizon to distant plasma, the BH’s rotational energy can be extracted via the Blandford-Znajek (BZ) mechanism \citep{1977MNRAS.179..433B}. This process drives a Poynting-flux-dominated jet, which can be self-collimated by the strong toroidal magnetic field generated by the rapidly rotating central engine \citep{2000PhR...325...83L, 2001PhR...345....1V}, with an estimated luminosity of 
\begin{equation}
L_{\rm BZ}=10^{50} \, a_\bullet^2 \left(\frac{m}{3}\right)^2\left(\frac{B}{10^{15} G}\right)^2 \ {\rm erg  \ s^{-1}},
\label{eq:Lbz} \end{equation}
\citep{1977MNRAS.179..433B}. By equating the magnetic pressure at the event horizon with the ram pressure of the innermost accretion flow \citep{1997rja..proc..110M}, the magnetic field strength threading the BH horizon is given by $B \simeq 2.3 \times 10^{16} \dot{m}_{-1}^{1/2} m^{-1} \left(1+\sqrt{1-a_\bullet^2} \right)^{-1} \rm{G}$.

\subsubsection{NS engine} \label{sec5_1_2} 

\begin{figure}[htbp]
    \centering
\includegraphics[height=6cm,width=8.5cm]{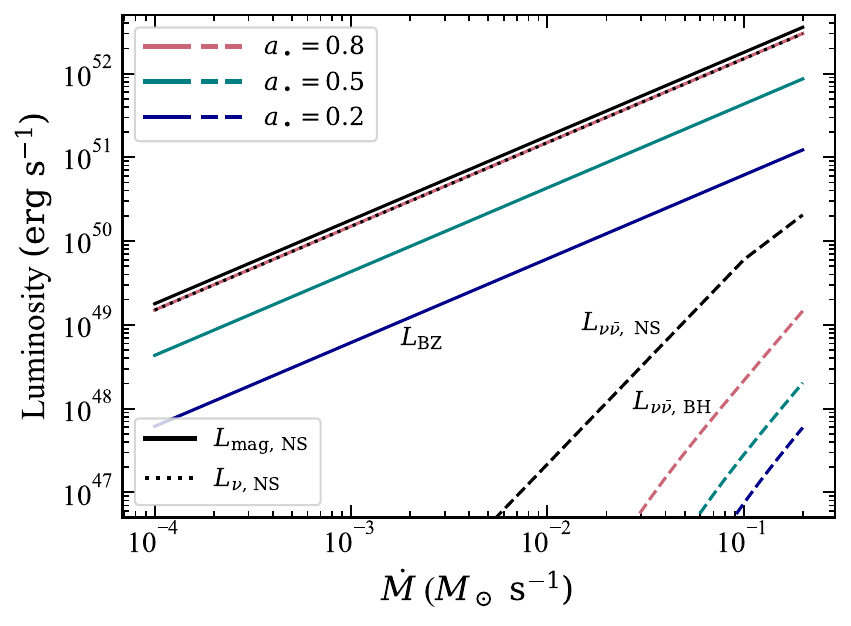}
    \caption{The dependence of various luminosities on the accretion rate. The colored lines correspond to BH engines, where the solid and dashed lines represent $L_{\rm BZ}$ and $L_{\nu\bar{\nu},\,\rm BH}$, respectively, calculated from Eqs.~(\ref{eq:Lbz}) and (\ref{eq:Lvv}) for $M= 10\, M_\odot$. The blue, green, and red colors correspond to $a_\bullet = 0.2$, $0.5$, and $0.8$, respectively. The black lines correspond to NS engines: $L_{\rm mag,\, NS}$ (solid) is calculated from Eq.~(\ref{eq:Lmag}), while $L_{\nu,\,\rm NS}$ (dotted) and $L_{\nu\bar{\nu},\,\rm NS}$ (dashed) are adopted from \cite[][their Figure~6]{2009ApJ...703..461Z}.}
    \label{Fig: dotm_lum}
\end{figure}

When the central engine is a NS, the stripped WD material may also form a neutrino-cooled accretion disk. Unlike a BH, a NS possesses a solid surface, which renders the innermost disk region denser and hotter. Consequently, the neutrino luminosity is significantly higher than that of a BH disk. \cite{2008ApJ...683..329Z, 2009ApJ...703..461Z, 2010ApJ...718..841Z} calculated the total neutrino emission luminosity $L_{\nu,\,\mathrm{NS}}$, including contributions from both the accretion disk and the NS surface, as well as the neutrino annihilation luminosity $L_{\nu\bar{\nu},\,\mathrm{NS}}$, as functions of the accretion rate $\dot{m}$, which we plot in Figure \ref{Fig: dotm_lum}.


On the other hand, if the accretion of the stripped WD material can significantly enhance both the spin rate and the surface magnetic field $B$ of a NS, a relativistic jet could also be produced via the magnetic dipole radiation. The corresponding luminosity is
\begin{equation} 
L_{\rm dip} \simeq 10^{49} B^2_{16}R^6_6 P^{-4}_{-3}\ {\rm erg\ s^{-1}},
\label{eq:Ldip} 
\end{equation} 
where $R=R_6 \times 10^6$ cm and $P= P_{-3}$ ms are the NS radius and spin period, respectively, and $B_{16}=B/10^{16}$ G.

The third mechanism for NS jet production is through magnetic dissipation. For an accreting NS, the angular momentum carried by the accreted material can induce differential rotation between the inner core and the outer envelope of the NS. The toroidal magnetic field can then be gradually amplified through magnetic field winding, forming a magnetically confined toroidal structure. Once the toroidal field strength exceeds the buoyancy threshold, the magnetic toroid can penetrate the NS surface and rapidly undergo magnetic reconnection, producing a magnetic bubble eruption \citep{1998ApJ...505L.113K, 2000ApJ...542..243R, 2006Sci...311.1127D, 2021NatAs...5..911Z, 2022Natur.612..232Y, 2023ApJ...947L..21Z}. Continuous accretion during the RPD phase may sustain this process and drive a relativistic jet, whose luminosity can be estimated as
\begin{equation}
L_{\rm mag, \, NS}=\eta \dot{M} c^2 \simeq  1.8\times 10^{51} \, \eta_{-1} \dot{m}_{-2} \ {\rm erg  \ s^{-1}},
\label{eq:Lmag}\end{equation}
where $\eta= 10\% \times \eta_{-1}$ is an umbrella efficiency factor for converting accretion power to jet power.

\subsection{Jet Property} \label{sec5_2}

Regardless of whether the central engine is a BH or a NS, the jet luminosity is sensitive to its spin and the accretion rate (\S \ref{sec5_1}). Figure \ref{Fig: dotm_lum} summarizes and shows various luminosities estimated in the above, for both BH and NS engines, in the range of accretion rates that are relevant to our simulation runs.  

However, due to their old age, one should not expect the pre-merger BH or WD to be rapidly rotating. Therefore, the change of spin $\Delta a_\bullet$ brought by the accretion during the RPD phase matters.
Our simulations found $\Delta a_\bullet \sim 0.06 - 0.86$ (see Table \ref{tab:tab1}).
In Category I, the central object is a relatively low-mass BH or NS, whereas the WD is heavy, resulting in a larger amount of accreted material. Consequently, the RPD phase sees a significant increase of $a_\bullet$. In contrast, in Categories II and III, the accreted mass is small compared with the BH mass. Thus, their $\Delta a_\bullet$ are small.
Therefore, wherever the central engine's spin is needed when estimating the jet property, we assign different fiducial values $a_\bullet =$ 0.8, 0.5 and 0.2 to Categories I, II and III systems, respectively.

In the following, we estimate the baryon loading rates and the terminal Lorentz factors (LFs) of the jets produced in the WD-NS/BH mergers, for both the BH and NS engines.

\subsubsection{BH-engine jet} \label{sec5_2_1} 

For jets produced from a BH engine, Figure \ref{Fig: dotm_lum} shows that $L_{\rm BZ}/L_{\nu\bar{\nu},\, \rm BH} \gtrsim 10^{3}$, suggesting that the jet launched is more likely by the BZ mechanism.

During the RPD of a WD by a BH, the formation and subsequent evolution of an accretion disk potentially involve complex changes in disk structure and geometry. When $\dot{m}>\dot{m}_{\rm ign}$, neutrino cooling becomes highly efficient, giving rise to a neutrino-dominated accretion flow. In this regime, neutrino-driven winds carry substantial baryonic matter and contaminate the jet, leading to a baryon-rich (``dirty'') fireball, which may significantly suppress the jet LF.

By contrast, in jets dominated by the BZ mechanism, magnetic fields threading the BH can establish a strong magnetic barrier that inhibits charged baryons (i.e., protons) from entering the jet, thereby producing a baryon-poor outflow. Nevertheless, neutrons entering the jet may still be converted into protons through various processes.

\cite{2013ApJ...765..125L,2017ApJ...849...47L} estimate the baryon-loading rate in BZ-powered jets as
\begin{multline}  \label{eq:dotm_bl}
\dot{M}_{\rm bl} \simeq 
3.5\times10^{-7} f_{\rm p,-1}^{-0.5}\theta_{\rm j,-1}\theta_{\rm B,-2}^{-1} \alpha_{-1}^{0.38} \\
\times \epsilon_{-1}^{0.83}\dot{m}_{-1}^{0.83}
(m/3)^{-0.55} r_{z,11}^{0.5} \, M_{\odot}\,{\rm s}^{-1},
\end{multline} 
for $\dot{m} > \dot{m}_{\rm ign}$, and it scales as $\dot{M}_{\rm bl} \propto \dot{m}^{1.9}$ for $\dot{m} \leq \dot{m}_{\rm ign}$. Here $f_{\rm p}= 0.1 \, f_{\rm p,-1}$ is the proton fraction, $\theta_{\rm j}= 0.1\, \theta_{\rm j,-1}$ is the jet half-opening angle, $\theta_B= 0.01\,\theta_{\rm B,-2}$ is the opening angle of the magnetic field lines, $\alpha= 0.1\,\alpha_{-1}$ is the dimensionless viscosity parameter, $\epsilon \equiv \dot{E}_\nu/(\dot{M}c^2) = 0.1\,\epsilon_{-1}$ is the neutrino radiation efficiency, and $r_z = r_{z,11}\times10^{11}\,{\rm cm}$  is the distance from the BH in the jet direction. 

The theoretically maximum LF that a jet can attain is $\Gamma_{\max} \simeq \mu_0 \equiv 1+ \eta + {L}_{\rm BZ}/(\dot{M}_{\rm bl} c^2)$, where $\eta \equiv L_{\nu\bar{\nu},\rm BH}  / (\dot{M}_{\rm bl} c^{2})$ denotes the terminal LF achievable through thermal acceleration. During the early acceleration phase, the jet is accelerated by a combination of thermal and magnetic processes to an initial LF of $\Gamma_0 \simeq \max \left( \eta, \mu_0^{1/3} \right)$. For an unsteady jet, further acceleration may occur via magnetic dissipation at larger radii, leading to a final LF that satisfies \citep{2013ApJ...765..125L,2017ApJ...849...47L} 
\begin{equation}
\Gamma_0 < \Gamma < \Gamma_{\max}.
\end{equation}

\begin{figure}[htbp]
    \centering
    \includegraphics[height=6cm,width=8.5cm]{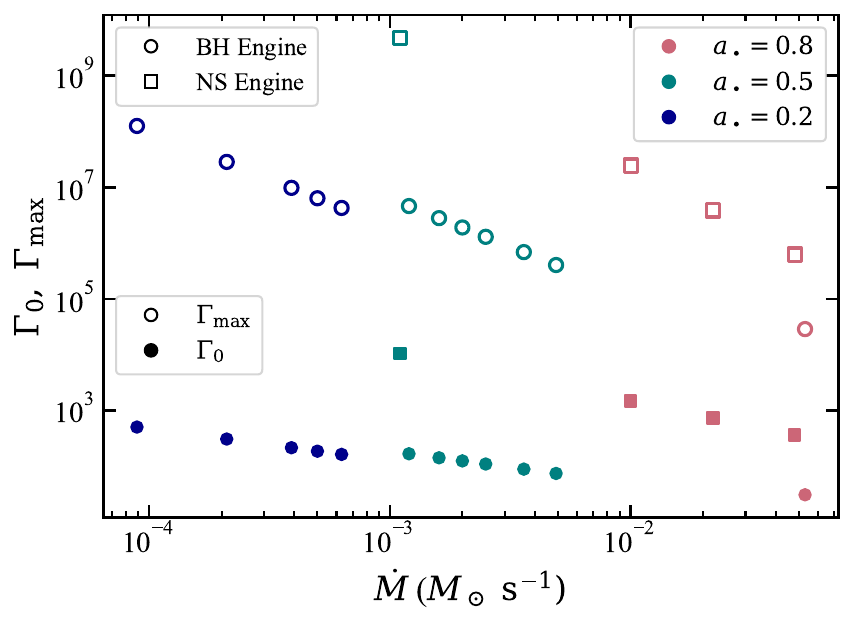}
    \caption{The expected asymptotic limits of jet LFs $\Gamma_0$ and $\Gamma_{\max}$, where Categories III, II and I are shown in blue, green and red colors, respectively; circles denote BH engines and squares denote NS engines. From Categories III to I, we assign ever increasing fiducial $a_\bullet$ values in the estimates, taking into account the increasing significance of the spinning up of the engines by the accretion ($\Delta a_\bullet$ in Table \ref{tab:tab1}) during RPDs.}
    \label{Fig: dotm_gamma}
\end{figure}

In Figure \ref{Fig: dotm_gamma} we plot the two asymptotic limits of jet LFs estimated from Eq. (\ref{eq:dotm_bl}) for all of our simulation runs, using each run's mean accretion rate and BH mass in Table \ref{tab:tab1} and adopting fiducial values for all other parameters. 
Note that these estimates are very idealized in nature and could be taken only by a grain of salt, since the jet acceleration and collimation involve many factors of uncertainties. 

As shown in Figure \ref{Fig: dotm_gamma}, at relatively low accretion rates, the baryon-loading rate in the BZ jet decreases rapidly with decreasing $\dot{M}$, resulting in a relatively clean jet. However, jets produced at low $\dot{M}$'s are also intrinsically less luminous (Figure \ref{Fig: dotm_lum}). At higher accretion rates, the jet luminosity is enhanced, but an excessively large $\dot{M}_{\mathrm{bl}}$ can suppress the jet LF.

The BH spin is another factor, as $L_{\rm BZ}$ is sensitive to $a_\bullet$. For lower-mass BHs (e.g., $M = 5 M_\odot$), $\Delta a_\bullet$ from the intense accretion during the RPD phase is substantial, thereby allowing the jet to achieve a relatively high LF, even if the pre-merger spin might be small.

If a relativistic jet can be launched, for $a_{\bullet} \sim 0.2 \text{-} 0.8 $ and $\dot{m} \sim 10^{-4} \text{-} 0.2 $, the luminosity of a BZ jet can reach $L_{\rm j,BZ} \sim 10^{47\text{-}53}\,\mathrm{erg\,s^{-1}} $ (see Figure \ref{Fig: dotm_lum}). For typical GRBs, the isotropic-equivalent energy lies in the range $E_{\rm iso} \sim 10^{49} \text{-} 10^{54}\,\mathrm{erg}$, whereas the true jet energy is confined to a narrow opening angle $\theta_{\rm j} \sim 0.05 \text{-} 0.2$ rad, such that $E_{\rm jet} = E_{\rm iso}({1-\cos\theta_{\rm j}})/{2} \sim 10^{46-52}\,\mathrm{erg}$.

Overall, in the WD-BH scenario, the BZ mechanism is more likely to launch a jet and is expected to produce predominantly X-ray transients. For lower-mass BHs, or for BHs with a sufficiently high pre-merger spin, the merger with a WD may, in principle, also produce a GRB.

\subsubsection{NS-engine jet} \label{sec5_2_2} 

For a NS engine, the energy losses are contributed from neutrino cooling, magnetic dipole radiation, and possibly accretion-induced magnetic reconnection events (\S \ref{sec5_1_2}). Figure~\ref{Fig: dotm_lum} plots the dependence of $L_{\nu,\,\rm NS}$,  $L_{\nu\bar{\nu},\,\rm NS}$ \citep{2009ApJ...703..461Z} and $L_{\rm mag,\, NS}$ (Eq.~\ref{eq:Lmag}) on the accretion rate. We find that if the accretion process triggers magnetic bubble eruptions, it has a better chance to dominate the jet power.

The neutrino heating on the NS surface could drive a substantial mass loss, with a rate estimated as \citep{1996ApJ...471..331Q}
\begin{multline}  \label{eq:mlost}
    \dot{M}_{\rm ml} \simeq 5.6 \times 10^{-11}  \, C^{10/3}  L_{{\nu},\rm NS,51} \\
   \times  \epsilon_{\nu,{\rm MeV}}^{10/3}  \,  R ^{5/3}_{6}  \, \left( \frac{m}{2} \right)^{-2} ~M_\odot~{\rm s^{-1}},
\end{multline}
where $\epsilon_{\nu, \,{\rm MeV}}=\epsilon_{\nu} /1~{\rm MeV}$ is the mean neutrino energy and $C \approx 1 +  
(m/2)/R_6$. It is the the mass-loss rate in the polar region that contributes to the baryon loading, whose rate can be estimated as $\dot{M}_{\rm po} = \dot{M}_{\rm ml}\, (\Delta\Omega/2\pi)\, f_{\rm asy},$ where $\Delta \Omega$ is the solid angle of the polar funnel and $f_{\rm asy}$ is a factor accounting for the mass loss' degree of deviation from spherical symmetry \citep{2009ApJ...703..461Z}.

The minimum bulk LF of the outflow emerging from the NS polar region can be estimated as
$\Gamma_{0} \simeq L_{\nu\bar{\nu},\,\rm NS} f_k / (\dot{M}_{\rm po} c^2)$,
where $f_k$ denotes the fraction of the energy deposited by neutrino-antineutrino annihilation that is converted into the kinetic energy of the NS outflow in the polar direction. If the RPDs significantly spin up the NS and accretion-induced magnetic reconnection events operate efficiently, the maximum bulk LF of the launched outflow can be estimated as $\Gamma_{\rm max} \simeq (L_{\rm mag,\,NS}+L_{\rm dip}+L_{\nu\bar{\nu},\,\rm NS} f_k)/(\dot{M}_{\rm po} c^2)$.

Adopting $f_{\rm asy} = f_k = 0.5$, $\Delta \Omega / 2\pi = 0.1$, spin period of 1 ms, and other parameter values appropriate for our simulation runs, we estimate $\Gamma_{\rm 0}$ and $\Gamma_{\rm max}$ for some of our simulation runs that are consistent with NS engines, and plot them as squares in Figure \ref{Fig: dotm_gamma}.

Figure \ref{Fig: dotm_gamma} shows that WD-NS systems may be capable of producing jets with high LFs. When considered together with the expected luminosities (see Figure  \ref{Fig: dotm_lum}) and beaming, WD-NS systems represent potential progenitors of X-ray transients or GRBs. However, NSs possess complex internal structures and magnetic field environments, and jet formation involves acceleration and dissipation processes that remain poorly understood \citep{2011MNRAS.413.2031M, 2017MNRAS.472.3058B, 2018ApJ...857...95M, 2011ApJ...726...90Z}. Therefore, the jet LFs predicted here should be regarded as highly idealized estimates.



\subsection{X-ray transients and GRBs} \label{sec5_3}

Below, we predict a variety of non-thermal transients from a jetted WD-BH/NS merger system that corresponds to Categories I, II, and III, respectively, of our simulation runs. They are: (i) X-ray transients with simultaneous GRBs; (ii) Long-duration X-ray transients accompanied by GRBs only at late times; (iii) Ultra-long X-ray transients.

It is important to note that the accretion rates derived from our simulations are not anticipated to directly mirror the realistic behavior of light curves. Since the observed emission depends on multiple factors, including accretion processes, jet dynamics, energy dissipation, and radiative processes (see discussion in \S\ref{sec4_3}, \S\ref{sec5_1} and \ref{sec5_2}), we do not model the realistic light curve and spectral evolution in this context. Instead, we just give some qualitative predictions on the possible observational signatures.

\begin{itemize}
\item X-ray transients with simultaneous GRBs
\end{itemize}

Category I generally corresponds to systems involving a more massive WD (e.g., $M_* \simeq 1.1$--$1.3\,M_\odot$) and a less massive (e.g., $\lesssim 5\, M_\odot$) BH or NS, resulting in an extremely violent RPD process and a very high accretion rate. We expect that Category I systems may produce both X-ray transients and GRBs simultaneously, with the duration of their prompt emission lasting up to $\sim 10^{2}$ s. 

Moreover, although the accretion rate in Category I shows large modulation amplitudes, the number of RPD cycles is relatively small. Consequently, the observed X-ray and $\gamma$-ray light curves may exhibit only a limited number of modulation periods, albeit with pronounced amplitudes.

\begin{itemize}
\item Long-duration X-ray transients accompanied by GRBs only at late times
\end{itemize}

The accretion history of Category II systems during the RPD process generally exhibits an increasing trend of $\dot{M}_{\rm peak}(t)$, reaching its maximum near the end of the RPD stage. Therefore, we expect that Category II systems may give rise to X-ray transients with durations of $10^{2-3}$ s, which could have a GRB appearing in the late phase of the X-ray transient. 

In addition, the number of accretion rate modulations in Category II events is larger than that in Category I, implying that the X-ray and $\gamma$-ray light curves of Category II events may contain more modulation cycles, which makes them easier to identify observationally. Moreover, the overall modulation amplitude is also relatively large.

\begin{itemize}
\item Ultra-long X-ray transients
\end{itemize}

Compared with Categories I and II, Category III generally corresponds to systems involving a less massive WD (e.g., $M_* \lesssim  1 \,M_\odot$) and a more massive BH (e.g., $M_* \gtrsim  10 \,M_\odot$). During the early stage of the RPD process, the accretion rate is relatively low and the jet is weak. As the accretion rate increases, the jet becomes stronger and then may produce an X-ray transient with a duration of up to $10^{3}$ s. It might lack a GRB because the jet is not sufficiently strong. 

The X-ray transient light curve would exhibit a larger number of modulation cycles, making it easier to identify observationally. However, in systems with lower orbital eccentricities, the modulation amplitude might be weak.

In summary, we expect that during the RPD phase, Categories I, II, and III may give rise to X-ray transients with different durations, with Categories I and II possibly accompanied by GRBs. In addition to the prompt emission, the subsequent deceleration of the jet or outflows is expected to give rise to afterglow emissions. Both the prompt and early afterglow emissions of these X-ray transients are good prospect sources for wide-field X-ray instruments, such as the Einstein Probe (EP, \citeauthor{2015arXiv150607735Y} \citeyear{2015arXiv150607735Y}).

\subsubsection{Possible AIC in WD-NS mergers} \label{sec5_1_3} 

Note that in scenarios involving a high-mass NS, as the mass of the NS grows up to a limit $\sim 3\ M_{\odot}$ during the RPD process, it would collapse into a BH, a phenomenon known as the accretion-induced collapse \cite[AIC, ][]{qin_collapse_1998,dermer_collapse_2006,perna_accretion-induced_2021}. For example, in Category I scenarios with an initial NS mass of $2\ M_{\odot}$, after the RPDs the NS can increase its mass to $2.4-2.7\ M_{\odot}$.

An intriguing electromagnetic signature may be associated with this collapse. During the AIC, the formation of an event horizon conceals most of the matter and radiation, with the exception of the NS magnetosphere: it undergoes violent disruption, leading to magnetic reconnection and the release of high-energy electromagnetic radiation. The exact characteristic of this radiation is still under debate, but it has been proposed as a potential explanation to both long or short GRBs \citep{vietri_supranova_1999,dermer_collapse_2006}, as well as fast radio bursts \citep{falcke_fast_2014}. 


One possible observational feature associated with the AIC, when it happens during a WD-NS merger, might be a sudden turn-off of the jet; if the accretion continues after the BH formation, the jet may resume, leaving a short gap in the jet emission light curve. Alternatively, if the AIC happens long after the cease of accretion, during the spinning down of the NS, it might cause a sharp drop-off to the spin-down energy injection power in the afterglow phase. Scenarios similar to the latter possibility have been discussed in the context of short GRBs \citep{2006Sci...311.1127D,gao_short-living_2006,metzger_short-duration_2008}.

The relative timing of the appearance (or the lack) of a possible AIC feature may help constrain the NS EOS, because the NS maximum stable mass depends on both the EOS and the rotation \citep{lyford_effects_2003}. This approach has been applied in modeling the X-ray plateau features seen in some short GRBs \citep[e.g.,][]{lu_millisecond_2015}. 


\subsection{Observational Cases} \label{sec5_5}  

GRBs 230307A and 211211A can be classified as Category I events in our WD-BH/NS scenario, based on their $\gamma$-ray emission properties. 
Based on the modulation periods of their prompt emission, \citetalias{2024ApJ...973L..33C} estimated the corresponding WD masses to be 1.3 and 1.4~$M_\odot$, respectively. These values are also consistent with the expected characteristics of Category I systems. Category I events are expected to exhibit nearly simultaneous GRBs and X-ray emission. For GRB 230307A, a soft X-ray burst was detected by LEIA almost simultaneously with the $\gamma$-ray emission \citep{2025NSRev..12E.401S}. For GRB 211211A, unfortunately, no X-ray instrument was monitoring the source during the prompt phase.

Another noteworthy feature of GRBs 230307A and 211211A is the presence of extended emission following the prompt phase. In Figure~\ref{Fig: accretion_rate}, the red dashed line marks the end of the RPD phase, corresponding to the time when the WD structure becomes almost completely destroyed in the simulation. As the main panel of Figure~\ref{Fig: binary_evolution} shows, at this point a final relic is left, whose accretion rate into the BH \textbf{or NS} evolves slowly, and the modulation feature nearly disappears. All of these are consistent with the observed characteristics of the extended emission from GRBs 230307A and 211211A.

Faint, rapidly evolving, and reddened transients powered by the radioactive decay of $^{56}$Ni in the ejecta of WD-BH/NS systems provide a plausible origin for the late-time kilonova-like optical/NIR transients observed in GRBs 230307A and 211211A \citep{2012MNRAS.419..827M, 2020MNRAS.493.3956Z, 2026A&A...706A.106K}. 

EP250702a / GRB 250702B is a remarkable transient that showed a $\sim$ 0.73-day (rest-frame) rise in soft X-rays, during which MeV emission was observed about three hours prior to the soft X-ray peak, resulting in at least three Fermi-GBM triggers \citep{2025arXiv250714286L, 2025arXiv250718694O,2025arXiv250925877L}. At a redshift of $z=1.036$, its isotropic-equivalent luminosity is
$L_{\gamma,\mathrm{iso}} = 5 \times 10^{51}\ \mathrm{erg\ s^{-1}}$ \citep{2025ApJ...994L..46C}.

Its prompt emission light curves in both MeV and soft X-ray bands show several distinct bursts. Recently, \cite{2025ApJ...994...21S} claimed possible quasi-periodic oscillation features in two of these bursts. This behavior is consistent with the RPD scenario of a WD-BH/NS merger and the prediction from our simulations. Moreover, the spectral evolution from soft X-ray to MeV energies, concurrent with the pre-peak rapid rise of the X-ray flux \citep{2025arXiv250925877L}, may indicate a progressively intensifying process, which is consistent with the ever-increasing trend of $\dot{M}_{\rm peak}(t)$ during the RPDs seen in our simulations. 

The fact that its MeV emission appeared only near the X-ray peak suggests a Category II classification in our WD-BH/NS scenario. However, the ultra-long duration ($\sim 6\times 10^4$ s) of the X-ray transient before the peak resembles the characteristic of Category III. Compared with the other two categories, producing a GRB in a Category III system is more challenging, although it remains possible when magnetic effects are taken into account. For EP250702a / GRB~250702B, assuming a jet half-opening angle in the range $\theta_j = 0.05\text{-}0.2$ rad, the inferred $\gamma$-ray luminosity is $L_\gamma \sim 10^{47\text{-}49}\,\mathrm{erg\ s^{-1}}$. As shown in Figure \ref{Fig: dotm_lum}, the WD-NS/BH scenario is a promising channel for producing a GRB with such luminosity. We note that a (or repeated) tidal disruption event of a WD by an intermediate-mass BH has been proposed to explain this transient \citep{2025arXiv250714286L, 2025arXiv250925877L, 2025arXiv250922843E, 2026arXiv260201073S, 2026arXiv260223299Y}. Overall, a WD-BH/NS system as the progenitor of EP250702a / GRB 250702B cannot be ruled out. 

\section{Conclusions} \label{sec6}

Recent studies suggest that mergers of WDs with NSs or stellar-mass BHs may be potential progenitors of unusually long GRBs, such as GRBs 230307A and 211211A. Compared to NS-NS or NS-BH mergers, the larger tidal disruption radius of a WD can result in a non-negligible residual orbital eccentricity during the late merger stage ($0 \lesssim e_0 \lesssim 0.2$). This residual eccentricity can lead to discontinuous mass transfer, with the WD undergoing RPDs. 

Using SPH simulations with $10^6$ particles, we modeled the late-stage mass transfer processes in WD-NS and WD-BH mergers, tracking the evolution of the accretion rate until accretion effectively ceased. Our main conclusions are as follows:

\begin{enumerate}

    \item We reproduce the process of a WD undergoing RPDs by a NS or BH. On an orbit with a small residual eccentricity, the WD experiences RPDs near $R_p$, which modulate the subsequent accretion process and cause the accretion rate to vary periodically with the orbital period.
    
    \item Simulations indicate that more compact configurations---with higher initial WD mass and lower BH (\textbf{or NS}) mass, smaller semi-major axis $a$, or shorter orbital period---exhibit higher expansion rates of the WD remnant ($|dR_*/dM_*|$) and higher $\dot{M}_{\rm mean}$. As a result, the RPD duration is shorter, and the number of cycles is reduced.
    
    \item In decreasing order of $\dot{M}_{mean}$, all simulation results can be classified into three categories.
\begin{itemize}
    \item Category I: Their $\dot{M}_{mean}$ exceeds $10^{-2}\,M_{\odot}\,\mathrm{s^{-1}}$, and the RPD phase can last up to $10^{2}$~s. Such RPD episodes occur less frequently but exhibit larger modulation amplitudes. They are likely to produce X-ray transients and GRBs almost simultaneously, with the prompt emission lasting up to $10^{2}$~s.
    
    \item Category II: Their $\dot{M}_{mean}$ lies in the range of $10^{-3}$--$10^{-2}\,M_{\odot}\,\mathrm{s^{-1}}$. The RPD phase typically lasts for $10^{2\text{-}3}$~s. These systems may give rise to long-duration X-ray transients lasting up to $10^{2\text{-}3}$~s, with GRB emission appearing only at later times. 

    \item Category III: Their $\dot{M}_{mean}$ is below $10^{-3}\,M_{\odot}\,\mathrm{s^{-1}}$. The duration of the RPD phase can reach $10^{3}$~s. Such systems may generate an ultra-long X-ray transient only (although the possibility of an accompanying GRB cannot be ruled out, it is less likely), with durations up to $10^{3}$~s.

\end{itemize}

    \item We anticipate that a common feature of these high-energy transients would be that their prompt emission---as opposed to their possible afterglow emissions---light curves are probably modulated to varying degrees by the orbital period.

    \item Our simulations further indicate that the prompt emissions of GRBs 230307A and 211211A originate from the RPDs in WD-BH/NS systems, while their extended emissions may be attributed to the accretion of the final debris after the WD's complete disruption. The recently detected EP250702a / GRB 250702B may also arise from this scenario. 
\end{enumerate}

\section*{Acknowledgments}
We thank Bing Zhang for valuable discussions and constructive comments and suggestions. JPC thanks Ming-Hao Zhang, Kun Liu, Rui-Qi Yang and Shang-Fei Liu for helpful discussions and technical support on the fluid simulation part. JPC thanks Qiang Wang, Ya-Cheng Kang, and Na Wang for carefully reading the manuscript posted on arXiv and for pointing out a few minor writing errors. We thank an anonymous reviewer for insightful comments and suggestions that greatly improved this work. This work is supported by National Key R\&D Program of China (Grant No. 2025YFF0511100) and by National Natural Science Foundation of China (No. 12393814 and 12503053). 
 
\bibliography{merger}{}
\bibliographystyle{aasjournalv7}

\end{CJK*}

\end{document}